\documentstyle[12pt]{article}
\begin{document}
\frenchspacing
\def\bbuildrel#1_#2^#3%
{\mathrel{\mathop{\kern 0pt#1}\limits_{#2}^{#3}}}
\newcommand{\trpsbs}[2]{_{{}_{\displaystyle #1_{{}_{\scriptstyle #2}}}}}

\mbox{ } \hfill{\normalsize MZ-TH/94-21}\\
\mbox{ } \hfill{\normalsize UCT-TP 214/94}\\
\mbox{ } \hfill{\normalsize August 1994\\

\begin{center}
{\Large \bf Mass singularities in light quark correlators:
the strange quark case}\\[1cm]

{\bf K.G. Chetyrkin}\\[.5cm]
Institute for Nuclear Research of the Russian Academy of Sciences,\\
60$^{th}$ October Anniversary Prospect 7a,\\
117312 Moscow, Russia\\[.75cm]
{\bf C.A. Dominguez}\footnote {John Simon Guggenheim
Fellow 1994-1995}\\[.5cm]
Institute of Theoretical Physics and Astrophysics\\
University of Cape Town, Rondebosch 7700, Cape, RSA\\[.75cm]
{\bf D. Pirjol}\footnote{Supported by the "Graduiertenkolleg Teilchenphysik",
University of Mainz}{\bf and K. Schilcher}\\[.5cm]
Institut f\"{u}r Physik,
Johannes Gutenberg-Universit\"{a}t\\
Staudingerweg 7, D-55099 Mainz, Germany\\
\end{center}
\vspace{0.5cm}
\begin{abstract}
\noindent
The correlators of light-quark currents contain mass-singularities of
the form $\log(m^2/Q^2)$. It has been known for quite some time that these
mass-logarithms can be absorbed into the vacuum expectation values of other
operators of appropriate dimension, provided that schemes without
normal-ordering are used. We discuss in detail this procedure
for the case of the mass logarithms $m^4\log(m^2/Q^2)$, including also the
mixing with the other dimension-4 operators to two-loop order. As an
application we present an improved QCD sum rule determination
of the strange-quark mass. We obtain $\bar m_s(1$ GeV$)=171\pm 15$ MeV.
\end{abstract}
\newpage
\section{Introduction}
The method of QCD sum rules, first introduced in \cite{SVZ}, has become
a popular and powerful technique to study QCD in the low-energy,
non-perturbative region. The starting point is the Operator Product
Expansion (OPE) of current correlators at short distances, suitably
modified to incorporate non-perturbative effects. The latter are
parametrized by a set of vacuum expectation values of the quark
and gluon fields entering the QCD Lagrangian. These vacuum condensates
induce power corrections to asymptotic freedom, and are responsible
for the rich resonance structure observed at low energies. The
basic assumption here is the factorization of short and long distance
effects. The former are associated with the Wilson coefficients in
the OPE, and the latter with the vacuum condensates. While the Wilson
coefficients are calculable in perturbation theory, to any desired
order in the strong coupling constant, the vacuum condensates cannot
be calculated analytically from first principles (this would be
tantamount to solving QCD exactly). Instead, they can be estimated
in the framework of lattice QCD, or extracted from experimental data
in certain channels by means of the QCD sum rules themselves. Next,
making use of the analyticity properties of the relevant Green functions,
and invoking the notion of QCD-hadron duality, one relates the
fundamental QCD parameters entering the OPE with a dispersive integral
involving the hadronic spectral function. In this fashion, a relation
between hadronic and QCD parameters is achieved.\\

An important problem which must be addressed in connection with
the factorization of short and long distance effects in the OPE is
the appearance of mass singularities in the coefficient functions.
They are actually a long-distance effect and thus their presence in the
coefficient functions spoils the desired factorization.
It is possible, however, to
shift them into the vacuum condensates,
provided one is willing to accept the existence of perturbative vacuum
expectation values of operators. This is equivalent to giving up the
customary normal-ordering prescription which, by definition,
sets such contributions to zero.
A detailed discussion of how this can be achieved is
presented in Section 2, including the renormalization group improvement.
As a phenomenological application of these results, we address in Section
3 the problem of determining the value of the strange-quark mass.
Some time ago \cite{CePa} a redetermination of the strange-quark mass
was performed in the framework of QCD sum rules, exploiting
new developments in the theoretical \cite{BroadGen87} and
experimental \cite{5} understanding of the two-point function involving
the strangeness-changing vector current divergence. This constituted an
improvement over earlier determinations of $m_{s}$ \cite{3}-\cite{DL}.
Of particular importance was the removal of logarithmic quark-mass
singularities in two-loop quark-mass corrections of order ${\cal O}
(m_{q}^{2})$ and ${\cal O}(m_{q}^{4})$, achieved in \cite{BroadGen87}.
However, a feature of some concern was the appearance of parametrically
enhanced terms of order ${\cal O} (1/\alpha_{s})$. In this paper we remedy
this problem, showing how these terms can be effectively avoided by
employing a scheme without normal ordering. Working at the three-loop
level in perturbative QCD, and including two-loop radiative corrections
to the condensates, we obtain a new expression for the current correlator.
This is then used in order to obtain an improved value of
the strange-quark mass.\\

\section{The Operator Product Expansion}
We will be concerned in the following with the vacuum expectation
value of the following time ordered product
\begin{equation}
T(q) =  i\int\mbox{\rm d} x \, \mbox{{\rm e}}
^{i q x }\, T(J(x) J^\dagger (0))\,,
\end{equation}
where $J = \partial_\alpha \overline s \gamma^\alpha u = m_s
\overline s i u$, and
the up- and down-quark masses are neglected.
Except for a sign change, which will
be given explicitly, all our results will hold also for the divergence
of the strangeness-changing axial vector current $J = \partial_\alpha
\overline s
\gamma^\alpha\gamma_5 u = m_s \overline s i\gamma_5 u$.
When sandwiched between vacuum states, the T-product (1) becomes
the corresponding 2-point correlator
\begin{equation}
 \psi(Q^2,\alpha_s,m,\mu) =  \langle 0| T(q)|0\rangle,
\end{equation}
where $\mu$ is the renormalization scale. Note that the polarization
operator is {\em not} renormalization-group~(RG)~invariant as the function
$\langle 0|T(J(x) J^\dagger (0))|0\rangle$ contains
non-integrable singularities in the vicinity of the point $x=0$.
These cannot be removed by the quark mass and coupling constant
renormalizations alone, but must be subtracted independently. We write the
correlator (2) as an expansion in powers of $1/Q$ as
\begin{eqnarray}
\psi(Q^2) &=& m_s^2\{\Pi_0(L,\alpha_s)Q^2 + m_s^2\Pi_2(L,\alpha_s)
\pm \frac{C_u(L,\alpha_s)}{Q^2}\langle m_u\bar uu\rangle_0\nonumber\\
&+& \sum_{j=1}^3\frac{C_j(L,\alpha_s)}{Q^2} \langle O_j\rangle
+ {\cal O}(Q^{-4})\}\,,
\end{eqnarray}
where $L=\ln(\mu^2/Q^2)$, and $Q^2=-q^2$. The upper sign in front of the
coefficient function $C_u$ corresponds to the scalar case and the lower one
to the pseudoscalar case.
The operators $O_i$ are
\begin{eqnarray}
O_1 = \frac{\alpha_s}{\pi}G_{\mu\nu}^aG^{a\mu\nu}\,,\quad
O_2 = m_s\bar ss\,,\quad
O_3 = m_s^4\,,
\end{eqnarray}
and the explicit expressions of the functions $\Pi_{0}$ and $\Pi_{2}$ will
be given later.
The only terms in $\psi$ which are not RG-invariant are the coefficient
functions $m_s^2\Pi_0$ and $m_s^4\Pi_2$. They satisfy the nonhomogeneous
RG equations
\begin{equation}
\mu\frac{d}{d\mu}(m_s^2\Pi_0) = m_s^2\gamma_q\,,\quad
\mu\frac{d}{d\mu}(m_s^4\Pi_2) = m_s^4\gamma_m\,,
\end{equation}
with
\begin{eqnarray}
\gamma_q &=& \frac{1}{8\pi^2}\left(-6-10\frac{\alpha_s}{\pi}
+ (3\zeta(3)-\frac{383}{12}) \frac{\alpha_s^2}{\pi^2}
   \right)\,,\\
\gamma_m &=& \frac{1}{8\pi^2}(-12-16\frac{\alpha_s}{\pi})\,.
\end{eqnarray}
The anomalous dimension $\gamma_q$ has been given to three-loop order
because the corresponding correction to $\Pi_0$ is a priori
not negligible, and it will be taken into account in the next Section.
The absorptive part of $\psi(Q^2)$, being an observable quantity,
is invariant under the RG transformations.
Without any loss of generality we will work with the second derivative
$\psi''(Q^2) \equiv d^2\psi(Q^2)/d(Q^2)^2$, which can be seen from (3)
and (5) to satisfy an homogeneous RG equation
\begin{equation}
\mu \frac{d}{d\mu}\psi''(Q^2) = 0\,.
\end{equation}
The high energy behavior of $\psi''(Q^2)$ in the deep euclidean region
may be
reliably evaluated in QCD by employing the operator product
expansion, i.e.
\begin{eqnarray}
& &Q^2\psi''(Q^2,\alpha_s,m_s,\mu)
\bbuildrel{=\!=\!\Longrightarrow}_{\scriptstyle{Q^2\to\infty}}^{}
K_0(Q^2,\alpha_s,m_s,\mu)\, \mbox{{\bf \large 1}}\nonumber\\
& & + \sum_{n}\frac{m_s^2}{(Q^2)^{n/2}}\sum_{dim\, O_i = n}
K_i( Q^2,\alpha_s,m_s,\mu) \langle 0| O_i (\mu)|0\rangle\,.
\end{eqnarray}
We have explicitly  separated the contribution  of the unit operator  from
that of the operators with a non trivial dependence on the field variables.
The coefficient functions (CF) $K_0$ and $K_i$  depend upon the
details of the renormalization prescription for the composite operators $O_i$.
The usual procedure of normal ordering for the composite
operators appearing on the r.h.s. of the OPE (9) becomes physically
unacceptable if  quark mass corrections are to be included.  This is
already obvious  for the unit operator, representing the usual
perturbative contributions if normal ordering is used,
because in general it contains  mass and momentum logarithms of the form
\begin{equation}
m_s^2(\frac{m_s^2}{Q^2})^n (\ln\frac{\mu^2}{Q^2})^{\alpha}
				(\ln\frac{\mu^2}{m_s^2})^{\beta},
\end{equation}
with $n, \, \alpha$ and $\beta$  being non-negative integers.
More specifically,  one may write \cite{CheSpi88}
\begin{equation}
K^{NO}_0(Q^2,\alpha_s,m_s,\mu)
\bbuildrel{=\!=\!\Longrightarrow}_{\scriptstyle{Q^2\to\infty}}^{}
m_s^2
\sum_{n \ge 0,\, l \ge 0}(\frac{m_s^2}{Q^2})^{n/2} (\frac{\alpha_s}{\pi})
^{l - 1}
F_{il}(L,M)\,,
\end{equation}
where
$L = \ln(\mu^2/Q^2)$,  $M = \ln(\mu^2/m_s^2)$, and
the overscript NO is a reminder of the normal ordering prescription
being used.
The function $F_{il}(L,M)$ corresponds to the contribution of the
$l$-loop diagrams, and is a polynomial of degree not higher than $l$,
in both $L$ and $M$. Now  it is obvious that one may not choose the
normalization scale $\mu$ in such a way that for $Q \gg m$ both $M$
and $L$ would be small. The mass logarithms signal that even in the
framework of perturbation theory there are effects coming from large
distances of order $1/m$.
Fortunately, it has been realized long ago \cite{BroadGen87},\cite{CheSpi88},
\cite{T83}-\cite{BroadGen84}
that  all the mass
logarithms may be neatly shifted into the vacuum expectation values
(VEV) of non-trivial composite operators appearing on the r.h.s. of
(9), provided the latter are minimally subtracted.

   To give a
simple example, let us consider the correlator (2) in the
lowest order one-loop approximation.
First, we use the normal ordering prescription for the composite operators
which appear into an OPE of the time ordered product in (1). To determine
the coefficients of the various operators, one possible method is to sandwich
both sides of the OPE between appropriate external
states. By choosing them  to be the vacuum, only the unit
operator {\bf 1} will contribute on the r.h.s., {\em if} the normal ordering
prescription is used. This means that the bare loop of Fig.1 contributes
entirely to the coefficient $K_0$ in (9). A simple calculation gives
\cite{Broad81} (in the sequel we neglect
all  terms of order $1/Q^6$ and higher ):
\begin{equation}
K_0^{NO}(Q^2,m_s,\mu) = Q^2\psi''(Q^2,m_s,\alpha_s,\mu)|
{{}_{{}_{\scriptstyle \alpha_s =0}}}
=\frac{3}{8\pi^2} m_s^2
\left[1 - 2\frac{m_s^2}{Q^2} - \frac{2m_s^4}{Q^4}(L - M)\right].
\end{equation}
This coefficient function contains mass-singularities (the M-term).
On the other hand, if one does not follow the normal ordering prescription,
then the operator $m_s\bar ss $ develops a non-trivial vacuum expectation
value even if the quark gluon interaction is turned off by
setting $\alpha_s = 0$.
Indeed, after minimally removing its pole singularity,
the one loop diagram of Fig. 2
leads to the following result \cite{Broad81}
\begin{equation}
\langle 0|\bar ss|0\rangle^{PT}=\frac{m_s^3}{16\pi^2}4N_c
\left(\ln \frac{\mu^2}{m_s^2}+1\right)\,.
\end{equation}
By inserting this into (9), the new coefficient function $K_0$ can be
extracted, with the result
\begin{equation}
K_0 = \frac{3m_s^2}{8\pi^2}
\left[1-2 \frac{m_s^2}{Q^2} - \frac{2m_s^4}{Q^4} (1 + L)\right].
\end{equation}
The mass logarithms are now completely transferred from the CF $K_0$ to
the VEV of
the quark operator (13)! The same phenomenon continues to hold even after
the $\alpha_s$ corrections are taken into account for (pseudo)scalar
and (pseudo)vector correlators, independently of their flavour
structure \cite{BroadGen87,Gen89}. The coefficient functions of the non-trivial
operators will also depend on whether or not normal ordering is employed.

   The underlying reason for this was first established in
\cite{me82b}. Here it was discovered that if the minimal
subtraction procedure is scrupulously observed \cite{ft1}
then no CF may depend on  mass logarithms in
every order of perturbation theory, irrespectively of the specific model,
and/or  OPE at hand.  This implies that all the mass logarithms
log $M$ in
(11) go over into the ``condensates'',  where they are hidden among
various non-perturbative contributions. This remarkable property leads to
the possibility of using the standard
RG techniques to study
mass effects in the framework of  QCD sum rules without interference
from unwanted mass singularities.
On the other hand, the above nice features of  minimal subtraction
come at a price:  when schemes without normal ordering are employed, then
the renormalization properties of composite operators and CF's
become more involved. This may already be observed in our one-loop
example. Indeed, as a consequence of (8) one can immediately infer
that the CF $K_0^{NO}$ is RG invariant and hence, should obey the
equation
\begin{equation}
\mu \frac{d}{d \mu}K_0^{NO}(Q^2)   = 0\,.
\end{equation}
This equation is satisfied  trivially  for (12) but {\em not} for $K_0$ as
expressed by (14)!
The reason is that the operator $m_s\bar ss$ ceases to be RG invariant
in the world without normal ordering. The vacuum diagram of Fig.2
has a divergent part which has to be removed by a new counterterm proportional
to the operator $m_s^4\mbox{{\bf \large 1}}$. In other words, $m_s\bar ss$
begins to mix with the ``operator'' $m_s^4\mbox{{\bf \large 1}}$
\cite{CheSpi88}.\\

To lowest order, the corresponding anomalous dimension matrix reads
\begin{equation}
\mu\frac{d}{d\mu}
\left(
\begin{array}{c}
m_s \bar ss \\
m_s^4
\end{array}\right)=
\left(\begin{array}{cc}
0 & \frac{3}{2\pi^2} \\
0 & -8\frac{\alpha_s}{\pi}
\end{array}\right)
\left(\begin{array}{c}
 m_s \bar ss  \\
 m_s^4
\end{array}\right)\,.
\end{equation}
The nonvanishing off-diagonal matrix element describes the mixing of the two
operators under renormalization and was obtained from the divergent part of
the vacuum diagram in Fig.2.
The diagonal matrix elements are just the anomalous dimensions of the
respective operators in the usual normal-ordering scheme. The lower one
is equal to $-4\gamma(\alpha_s)$, where $\gamma(\alpha_s)$ is the strange
quark mass anomalous dimension which defines its running according to
\begin{equation}
\mu\frac{\mbox{d}}{\mbox{d}\mu}m_s = -\gamma(\alpha_s) m_s\,,
\end{equation}
where \cite{Tar82}
\begin{equation}
\gamma(\alpha_s)=\gamma_1\frac{\alpha_s}{\pi} +
\gamma_2\left(\frac{\alpha_s}{\pi}\right)^2 + \cdots\,,
\end{equation}
with
$\gamma_1 = 2$,
\begin{eqnarray}
\gamma_2 &=& \frac{101}{12}-\frac{5}{18}n_f\,,\\
\gamma_3 &=&  \left(1249 -
\left[\frac{2216}{27}+\frac{160}{3}\zeta(3)\right]
n_f-\frac{140}{81}n_f^2\right)/32 \,,
\end{eqnarray}
and $n_f$ is the number of  active light quarks.
The running of the coupling constant $\alpha_s(\mu)$ is determined by
\begin{equation}
\mu\frac{\mbox{d}}{\mbox{d}\mu}\alpha_s = \beta(\alpha_s)\alpha_s\,,
\end{equation}
where \cite{TarVlaZha80}
\begin{equation}
\beta(\alpha_s)=\beta_1\frac{\alpha_s}{\pi} +
\beta_2\left(\frac{\alpha_s}{\pi}\right)^2 +
\beta_3\left(\frac{\alpha_s}{\pi}\right)^3 + \cdots\,,
\end{equation}
with
\begin{eqnarray}
\beta_1 &=& -\frac{11}{2}+\frac{1}{3}n_f\,,
\nonumber\\
\beta_2 &=& -\frac{51}{4}
+ \frac{19}{12}n_f\, ,
\nonumber\\
\beta_3 &=&  \left(-\frac{2857}{2} + \frac{5033}{18}n_f
- \frac{325}{54}n_f^2\right)/32\,.
\end{eqnarray}
The solutions of (17) and (21) can be written as
\begin{eqnarray}
\alpha_s(\mu) = \frac{2\pi}{-\beta_1 L}
\left(
1 + \frac{2\beta_2}{\beta_1^2}\frac{\ln L}{L}
  + \frac{4}{\beta_1^2 L^2}
\left(
\frac{\beta_2^2}{\beta_1^2}
(\ln^2 L - \ln L - 1) + \frac{\beta_3}{\beta_1}
\right)
\right)\,,
\end{eqnarray}
\begin{eqnarray}
&{}& {m_s(\mu)}  =
\frac{\hat m_s}{(\frac{1}{2} L)^{-\gamma_1/\beta_1}}
\left(1 - \frac{2\gamma_1\beta_2}{\beta_1^3}
\frac{LL}{L}
+
\frac{2}{\beta_1^2}\left(\gamma_2-\frac{\gamma_1\beta_2}{\beta_1}\right)
\frac{1}{L}
\right.
\nonumber
\\
&{}&
+
\frac{2}{\beta_1^6 L^2}
( -\beta_1^3\gamma_3
  + \beta_1^2\gamma_2^2
  +  \beta_1^2\beta_2\gamma_2
  +  \beta_1\beta_2^2\gamma_1
  -  \beta_1^2\beta_3\gamma_1
  -  2\beta_1\beta_2\gamma_1\gamma_2
  +  \beta_2^2\gamma_1^2
)
\nonumber
\\
&{}&
+
\left.
\frac{4\ln L}{\beta_1^6 L^2}
( \beta_1^2\beta_2\gamma_2
  - \beta_1\beta_2\gamma_1\gamma_2
  +  \beta_2^2\gamma_1^2
)
+
\frac{2\ln^2 L}{\beta_1^6 L^2}
( - \beta_1\beta_2^2\gamma_1
  + \beta_2^2\gamma_1^2
)
\right)
\,.
\end{eqnarray}
The expressions (24) and (25) are given to three-loop order
for completeness. The simplified arguments so far will only
make use of the corresponding one-loop results.
Now one can see that the operator $m_s\bar ss$ acquires a scale
dependence, which can be obtained from (16) and is given by \cite{CheSpi88}
\begin{eqnarray}
m_s\bar ss(\mu) &=& m_s\bar ss(\mu_0) + \frac{3}{2\pi^2}
\int_{\alpha_s(\mu_0)}^{\alpha_s(\mu)}\frac{\mbox{d}x}{x\beta(x)}
\exp\left(-4\int_{\alpha_s(\mu_0)}^x\frac{\mbox{d}y}{y}\frac{\gamma(y)}
{\beta(y)}\right)m_s^4(\mu_0)\nonumber\\
&=& m_s\bar ss(\mu_0)
-\frac{3}{2\pi(4\gamma_1+\beta_1)}\left(\frac{m_s^4(\mu)}{\alpha_s(\mu)}
-\frac{m_s^4(\mu_0)}{\alpha_s(\mu_0)}\right)\,.
\end{eqnarray}
A distinctive feature of this result is the appearance of inverse powers of
$\alpha_s$ \cite{BroadGen84}.
Note that in the approximation we have considered, the combination
\begin{equation}
I_s = (m_s\bar ss)(\mu) + \frac{3}{2\pi(4\gamma_1+\beta_1)}
\frac{m_s^4(\mu)}{\alpha_s(\mu)}
\end{equation}
is a RG invariant. It corresponds (but is not generally equal) to the
RG-invariant combination $m_s\bar ss$ in the usual normal-ordering scheme.
For simplicity, we will neglect for the moment the contributions of the
dimension-4 operators $G^2$ and $\bar uu$ in the OPE (9). They will be
added in later. The lowest-order coefficients of the operators
$m_s\bar ss$ and $m_s^4$ at the scale $\mu=Q$ have the values:
\begin{equation}
c_{m_s\bar ss} = \frac{1}{2}\,,\quad  c_{m_s^4} = \frac{3}{16\pi^2}\,.
\end{equation}
We are now in a position to derive the RG improvement
of the coefficient functions appearing in the OPE (9) when working in a
scheme without normal ordering. To achieve this, one
notes from Eq.(8) that the total contribution of the operators
of dimension 4 is RG invariant and therefore we can choose freely the scale
$\mu$.
Setting the renormalization scale $\mu=Q$ allows us to absorb all
logarithms $\ln \mu^2/Q^2$ appearing in the CF $K_0$ into the running
coupling constant and the strange quark mass. On the other hand, the matrix
elements of the
operators at this scale can be expressed in terms of the same matrix
elements at a lower scale $\mu_0\simeq 1$ GeV$^2$ with the help of the
RG equation (16).
Our first result for the RG improvement of the OPE (9), treated entirely
within the minimal subtraction prescription, reads (for $n_f=3$)
\begin{eqnarray}
& & Q^2\psi''(Q^2,m_s,\mu)
\bbuildrel{=\!=\!\Longrightarrow}_{\scriptstyle{Q^2\to\infty}}^{}
\frac{3}{8\pi^2}m_s^2(Q)
\left(1 - 2 \frac{m_s^2(Q)}{Q^2}\right)\nonumber\\
&+& \frac{m_s^2(Q)}{Q^4}\left(
\langle 0|(m_s\bar ss)(\mu_0)|0\rangle - \frac{3}{7\pi}\frac{m_s^4(Q)}
{\alpha_s(Q)} + \frac{3}{7\pi}\frac{m_s^4(\mu_0)}{\alpha_s(\mu_0)}\right)
+ O(1/Q^6) \,.
\end{eqnarray}
This result has been essentially obtained for the first time in
\cite{BroadGen84} (to two loop order). There are, however, a number of
differences between its
interpretation as given in \cite{BroadGen84} (see also \cite{BroadGen87}
and \cite{Gen89})
and the point of view we will take in this paper.
We comment briefly on these differences.
In \cite{BroadGen84} the vacuum expectation value of the
RG invariant combination $I_s$ in (27) was
identified with the (RG-invariant) product $m_s\langle 0|\bar ss^{NO}|
0\rangle$ in the usual scheme using normal ordering. Thus, the vacuum
matrix element $\langle 0|(m_s\bar ss)(\mu)|0\rangle$ (non-normal ordered
and RG-noninvariant) was represented, to the order we are working, as the sum
of a RG-invariant part (of a nonperturbative origin, due to the spontaneous
breaking of the chiral symmetry in QCD) and of a perturbative part
($\mu$-dependent) which represents the sum of the leading mass singularities
of the form $\alpha_s^n\ln^{n+1} (\mu^2/m^2)\,$:
\begin{equation}
\langle 0|(m_s\bar ss)(\mu)|0\rangle = \langle 0|m_s\bar ss^{NO}|0\rangle
-\frac{3}{7\pi}\frac{m_s^4(\mu)}{\alpha_s(\mu)}\,.
\end{equation}
As it can be seen from (24) and (25) in the one-loop
approximation, the second term vanishes in the limit $\mu\to\infty$
and the distinction between the VEV of the operator $m_s\bar ss$ in
schemes with and without normal ordering disappears.
It should be stressed that the above interpretation of the relation
between normal ordered and non-ordered  quark condensates
relies heavily on an implicit assumption which is difficult to
(dis)prove. Indeed, all purely perturbative contributions to
$\langle 0|(m_s\bar ss)(\mu)|0\rangle $ were assumed to vanish in the
limit $\mu \to \infty$. Fortunately, even if the hypothesis fails it
will only spoil the applicability of the scheme with normal ordering,
but would have no effect on  other renormalization schemes like e.g. the
minimal subtraction prescription.
The practical consequence of this approach is a large value of the
mass correction of order $m_s^4$, which is enhanced by the presence of
one negative power of $\alpha_s(Q)$ (the second term in the second line of
Eq.(29)). Note that in this approach there is no corresponding term
containing $1/\alpha_s(\mu_0)$ in Eq.(29), because it can be
effectively combined with the contribution of the operator $m_s\bar ss$
resulting in the RG-invariant VEV $\langle 0|I_s|0\rangle$.
On the other hand, considering that a typical momentum transfer for
QCD sum rules is of about 1 GeV, we will work with the quark and gluon
condensates normalized at this ``natural'' scale $\mu_0 = 1$ GeV as our
reference values.
(For the case of the semihadronic decay rate of the
$tau$ lepton a similar approach has been suggested
in \cite{Chetyrkin93}.)
As mentioned above, this point of view is equivalent to
the one taken in \cite{BroadGen84}, provided the scale $\mu_0$ would have
been taken to infinity. Our choice of $\mu_0$ somewhere around the
characteristic momentum scale specific to the problem at hand ($\simeq
1$ GeV) helps to avoid the parametrically enhanced inverse powers of
$\alpha_s$. Indeed, as one can see from (24) to one-loop order, one
has $\pi/\alpha_s(Q)-\pi/\alpha_s(\mu_0) = -\beta_1\ln(Q/\mu_0)$, which is
not particularly large.\\

\section{Determination of the Strange Quark Mass}
We proceed now to include the contributions from the gluon
operator $G^2$ and from the light quark condensate $\bar uu$, working
consistently to two-loop order.
The coefficient functions $\Pi_0$ and
$\Pi_2$ in (3) have, respectively, the three-loop \cite{GorKatLarSur8990}
and two-loop values (for an arbitrary
renormalization scale $\mu$, the use of the $\overline{MS}$ scheme is
understood)
\begin{eqnarray}
&{}&
\Pi_0 = \frac{1}{16\pi^2}
\left[
-12-6\ln\frac{\mu^2}{Q^2}
+ \frac{\alpha_s}{\pi}
     \left(
     -\frac{131}{2}
     -34\ln\frac{\mu^2}{Q^2} - 6\ln^2\frac{\mu^2}{Q^2} + 24\zeta(3)
     \right)
\right.
\nonumber
\\
&{}&
  +\,  (\frac{\alpha_s}{\pi})^2
\left(
	    - \frac{ 17645 }{ 24 }
	    + 353 \zeta(3)
	    - 8 n_f \zeta(3)
	    + \frac{ 511 }{ 18 } n_f
	    + \frac{ 3 }{ 2 } \zeta(4)
	    - 50 \zeta(5)
\right.
\nonumber
\\
&{}&
	    - 4 n_f \zeta(3)\ln\frac{\mu^2}{Q^2}
	    +  n_f\frac{ 65 }{ 4 } \ln\frac{\mu^2}{Q^2}
	    + 117\zeta(3) \ln\frac{\mu^2}{Q^2}
	    - \frac{ 10801 }{ 24 } \ln\frac{\mu^2}{Q^2}
\nonumber
\\
&{}&
\left.
\left.
	    + n_f \frac{ 11 }{ 3 } \ln^2\frac{\mu^2}{Q^2}
	    - 106 \ln^2\frac{\mu^2}{Q^2}
	    + n_f \frac{ 1 }{ 3 } \ln^3\frac{\mu^2}{Q^2}
	    - \frac{ 19 }{ 2 } \ln^3\frac{\mu^2}{Q^2}
\right)
\right] \,,
\end{eqnarray}
\begin{equation}
\Pi_2 = \frac{1}{16\pi^2}\left[
 -12-12\ln\frac{\mu^2}{Q^2} + \frac{\alpha_s}{\pi}\left(-100 -
64\ln\frac{\mu^2}{Q^2} - 24\ln^2\frac{\mu^2}{Q^2} + 48\zeta(3)\right)
\right]\,.
\end{equation}

   The three-loop result (31) was obtained in the works
\cite{GorKatLarSur8990} while
the $\alpha_s^2$ terms in (31,32) can be found in \cite{Broad81,Becchi81}.
Note that the first evaluation of $\Pi_0$ at three-loop level made in
Ref. \cite{Gorishny84a}
proved to be erroneous \cite{ft2}. Unfortunately, that wrong result was then
used in the work \cite{Gorishny84b} to find the light quark masses in the
framework of the finite energy sum rules.

The coefficient functions of the dimension-4 operators, evaluated at
the scale $\mu=Q$, are

\begin{eqnarray}
C_1 &=& \frac{1}{8}\left( 1+\frac{11\alpha_s}{2\pi}\right)\,,\\
C_2 &=& \frac{1}{2}\left( 1+\frac{11\alpha_s}{3\pi}\right)\,,\\
C_u &=& (1+\frac{14\alpha_s}{3\pi})\,,\\
C_3 &=& \frac{3}{16\pi^2}\left(1 + \frac{\alpha_s}{\pi}(8\zeta(3)-6)\right)\,.
\end{eqnarray}

   The leading order contributions to (33-35) were computed in \cite{SVZ}.
The two-loop corrections to (33-36) were evaluated in the Refs.
\cite{SurTka90}, \cite{SVZ,Becchi81}, \cite{SVZ,Becchi81} and
\cite{BroadGen87,Gen89} respectively.

   It is now a simple matter to derive the RG improvement
of the $\Pi_{0,2}$-terms in (3). Solving (5)
with the boundary conditions (31)-(32) yields
\begin{equation}
\begin{array}{c}
\displaystyle
{}
(m_s^2\Pi_0)|_{\mu}=
\\
\displaystyle
-\frac{3}{4\pi(2\gamma_1+\beta_1)}\frac{m_s^2(Q)}
{\alpha_s(Q)}
\left(1+(r_1+4+\beta_1)\frac{\alpha_s(Q)}{\pi}\right.\\
\displaystyle
 {}
+\left. (
r_3 + \frac{131}{6} + \frac{131\beta_1}{24} - 8\zeta(3) - 2\beta_1\zeta(3)
)
\left(\frac{\alpha_s(Q)}{\pi}\right)^2
\right)\\
\displaystyle
 {}
+\frac{3}{4\pi(2\gamma_1+\beta_1)}\frac{m_s^2(\mu)}{\alpha_s(\mu)}
\left(1+r_1\frac{\alpha_s(\mu)}{\pi}
+
r_3
\left(\frac{\alpha_s(\mu)}{\pi}\right)^2
\right)
\end{array}
\end{equation}
\begin{eqnarray}
(m_s^4\Pi_2)|_{\mu}&=&-\frac{3}{2\pi(4\gamma_1+\beta_1)}
\frac{m_s^4(Q)}{\alpha_s(Q)}
\left(1+(r_2+4+\frac{\beta_1}{2})\frac{\alpha_s(Q)}{\pi}\right)\nonumber\\
&+&\frac{3}{2\pi(4\gamma_1+\beta_1)}\frac{m_s^4(\mu)}{\alpha_s(\mu)}
\left(1+r_2\frac{\alpha_s(\mu)}{\pi}\right) \,,
\end{eqnarray}
where
\begin{equation}
r_1=\frac53 - \frac{\gamma_2}{2} + \frac{5\beta_1}{12} -
\frac{\beta_2}{4} = -2\,,\quad \mbox{ ( $n_f=3$ )}
\end{equation}
\begin{equation}
r_2=\frac43 - \frac{\gamma_2}{2} + \frac{\beta_1}{6} -
\frac{\beta_2}{8}\\
= -\frac{53}{24}\,,\quad \mbox{ ( $n_f=3$ )}
\end{equation}
\begin{eqnarray}
r_3 &=& \frac{9(7889-432\zeta(3))+9(439-1824\zeta(3))n_f-1195n_f^2}
{864(-57+2n_f)}
\nonumber
\\
&=& \frac{5904\zeta(3)-8011}{4896}\mbox{ ($n_f=3$)}\,.
\end{eqnarray}
The RG improvement of the contribution of the
dimension-4 operators in (3) requires the knowledge of their mixing
under renormalization. The generalization of (16) to two-loop order, by
taking also into account the mixing with the gluon operator $G^2$, reads
\cite{CheSpi88}
\begin{equation}
\mu\frac{d}{d\mu}
\left(
\begin{array}{c}
G^2 \\
m_s\bar ss \\
m_s^4
\end{array}\right)=
\left(\begin{array}{ccc}
-\alpha_s\frac{d\beta}{d\alpha_s} & -4\alpha_s\frac{d\gamma}{d\alpha_s} &
   4\alpha_s\frac{d\gamma_0}{d\alpha_s} \\
0 & 0 & -4\gamma_0 \\
0 & 0 & -4\gamma
\end{array}\right)
\left(\begin{array}{c}
 G^2 \\
 m_s \bar ss  \\
 m_s^4
\end{array}\right)\,.
\end{equation}
Here $\beta$ and $\gamma$ were defined in (21) and (17)
respectively, and $\gamma_0$ is the two-loop vacuum energy anomalous
dimension, given by
\begin{equation}
\gamma_0 = -\frac{3}{8\pi^2}\left(1+\frac{4}{3\pi}\alpha_s\right)\,.
\end{equation}
The operator $m_s\bar uu$ is RG invariant. The
analogous anomalous dimension matrix which describes the mixing of the
dimension-5 operators in schemes without normal ordering has been
calculated recently, to one-loop order, in \cite{JM}.
We apply now the RG improvement of the contribution
of the dimension-4 operators in (3). By taking advantage of the
fact that their total contribution is RG-invariant, we choose $\mu=Q$,
where the coefficient functions are given by (33)-(36). The
matrix elements of the operators $O_{1,2,3}$ can be scaled at $\mu_0\simeq
1$ GeV with the help of (42), where they are known. This procedure
leads to
\begin{eqnarray*}
& &\sum_{j=1}^3C_jO_j = (1+\frac{14\alpha_s(Q)}{3\pi})
   \langle m_s\bar uu\rangle_0
+\frac{1}{8}\left[ 1+\left(\frac{11}{2}-\frac{\beta_2}{\beta_1}\right)
\frac{\alpha_s(Q)}{\pi} \right.
\end{eqnarray*}
\begin{eqnarray*}
&+& \left. \frac{\beta_{2} \alpha_{s}(\mu_{0})}{\beta_{1}\pi}\right]
\langle O_1\rangle |_{\mu_0}
+\left[\frac{1}{2}+\left(\frac{11}{6}-\frac{\gamma_1}{2\beta_1}\right)
\frac{\alpha_s(Q)}{\pi}+\frac{\gamma_1}{2\beta_1}\frac{\alpha_s(\mu_0)}{\pi}
\right] \langle O_2\rangle |_{\mu_0}
\end{eqnarray*}
\begin{eqnarray*}
&-& \frac{3}{4\pi(4\gamma_1+\beta_1)\alpha_s(Q)}
\left(1+\frac{\alpha_s(Q)}{\pi}[r_2+\frac{47}{12}-\gamma_1-
\frac{\beta_1}{4}]\right)m_s^4(Q)
\end{eqnarray*}
\begin{eqnarray*}
&+& \frac{3}{4\pi(4\gamma_1+\beta_1)\alpha_s(\mu_0)}
\left(1+\frac{\alpha_s(\mu_0)}{\pi}[r_2+\frac{\gamma_1}{\beta_1}+
\frac{1}{4}]\right)m_s^4(\mu_0)
\end{eqnarray*}
\begin{eqnarray}
&+& \frac{1}{4\pi^2(4\gamma_1+\beta_1)}\frac{\alpha_s(Q)}{\alpha_s(\mu_0)}
m_s^4(\mu_0)\left(11-\frac{3\gamma_1}{\beta_1}\right) \,.
\end{eqnarray}
In order to keep the expressions within a
reasonable size, we will replace here the various constants by their
numerical values corresponding to $n_f=3$. At the same time, to help the
reader who might want to reproduce our result, the $n_f$-dependence which
appears from other sources  will be left explicit in the following.
Thus, putting together (37) and (44) and taking two
derivatives with respect to $Q^2$ we obtain
\begin{eqnarray}
\psi''(Q^2)&=&\frac{3m_s^2(Q)}{8\pi^2 Q^2} [
1+\frac{11\alpha_s(Q)}{3\pi}
+
(\frac{5071}{144} - \frac{35}{2}\zeta(3) )
\left(\frac{\alpha_s(Q)}{\pi}\right)^2
\nonumber
\\
&-&
2\frac{m_s^2(Q)}{Q^2}\left(1+\frac{28\alpha_s(Q)}{3\pi}\right) ]
\nonumber
\\
&+& \frac{m_s^2(Q)}{Q^6}\left\{
  2\langle m_s\bar uu\rangle
\left(1+\frac{23\alpha_s(Q)}{3\pi}\right)\right.
\nonumber
\\
&+& \frac{1}{4}\left.\langle\frac{\alpha_s}{\pi}G^2\rangle |_{\mu_0}
  \left(1+\frac{16\alpha_s(\mu_0)}{9\pi}
  +\frac{121\alpha_s(Q)}{18\pi}\right)\right.
\nonumber
\\
&+& \left.\langle m_s\bar ss\rangle |_{\mu_0}
  \left(1-\frac{4\alpha_s(\mu_0)}{9\pi}
  +\frac{64\alpha_s(Q)}{9\pi}\right)
- \frac{3}{7\pi^2}m_s^4(Q)
  \left(\frac{\pi}{\alpha_s(Q)}+\frac{155}{24}\right)\right.
\nonumber
\\
&+& \left.\frac{3}{7\pi^2}m_s^4(\mu_0)
 \left(\frac{\pi}{\alpha_s(\mu_0)}-\frac{173}{72}\right)
+\frac{64}{21\pi^2}\frac{\alpha_s(Q)}{\alpha_s(\mu_0)} m_s^4(\mu_0)
 \right\}\,.
\end{eqnarray}
A similar relation has been previously used  in a QCD sum rule determination
of the strange quark mass \cite{CePa}, where it was interpreted in the
spirit of \cite{BroadGen84}. As explained  earlier, in the
approach of \cite{BroadGen84} the normal-ordered strange quark condensate
(times $m_s$) is identified with the VEV of the RG-invariant combination
$I_s$ defined at one-loop level in Eq.(27). At two-loop level it has
the form
\begin{equation}
I_s = (m_s\bar ss)(\mu) + \frac{3}{2\pi(4\gamma_1+\beta_1)}
\frac{m_s^4(\mu)}{\alpha_s(\mu)}\left(1+r_2\frac{\alpha_s(\mu)}{\pi}\right)\,.
\end{equation}
Besides this, our result (45) differs from the one in \cite{BroadGen84}
(see also \cite{Gen89}) because there the mixing of the gluon condensate with
the other operators of dimension 4 has been neglected.\\

We perform the Borel transform $\hat L$ of $\psi''(Q^2)$, i.e.
\begin{eqnarray}
\hat L[\psi''(Q^2)] &=& \lim_{N\to\infty}\frac{(-1)^N}{(N-
1)!}(Q^2)^N
\frac{\partial^N}{(\partial Q^2)^N}\psi''(Q^2) \nonumber\\&=&
 \frac{1}{M^6}\int_0^\infty\mbox{d}s\,
  e^{-s/M^2}\frac{1}{\pi}\mbox{Im}\,\psi(s)\,.
\end{eqnarray}
A simple calculation using the methods of \cite{deRa} gives
\begin{eqnarray}
& &\hat L[\psi''(Q^2)] = \frac{3}{8\pi^2}\frac{\hat m_s^2}{M^2}
\frac{1}{[\frac12\ln (M^2/\Lambda^2)]^{-2\gamma_1/\beta_1}}
\nonumber
\\
&\times&\left( 1+\frac{4}{9\ln\frac{M^2}{\Lambda^2}}
\left[ \frac{11}{3}-\gamma_1\psi(1)
+\frac{4\beta_2}{\beta_1^2}\ln\ln\frac{M^2}{\Lambda^2} - \frac{4}{\beta_1
\gamma_1}\left(\gamma_2-\gamma_1\frac{\beta_2}{\beta_1}\right)\right]
\right.
\nonumber\\
&+&\left.\frac{2}{81\ln^2 (M^2/\Lambda^2)}\left\{\frac{2510167}{6561}-
\frac{1340}{9}\zeta(3)+34\psi^2(1)-\frac{17332}{81}\psi(1)-\frac{17}{3}
\pi^2\right.\right.\nonumber\\
&+&\left.\left.\ln\ln (M^2/\Lambda^2)\left(-\frac{1109248}{6561}+
\frac{4352}{81}\psi(1)\right) + \ln^2\ln (M^2/\Lambda^2)\frac{139264}{6561}
\right\}\right.\nonumber\\
&-&\left.2\frac{\hat m_s^2}{M^2}
\frac{1}{[\frac12\ln (M^2/\Lambda^2)]^{-2\gamma_1/\beta_1}}
\left\{1 + \frac{4}{9\ln\frac{M^2}{\Lambda^2}}
\left[\frac{28}{3}-2\gamma_1\psi(2)
+8\frac{\beta_2}{\beta_1^2}\ln\ln\frac{M^2}{\Lambda^2}\right.\right.\right.
\nonumber\\
&-&\left.\left.\left.\frac{8}{\beta_1\gamma_1}\left(\gamma_2-\gamma_1
\frac{\beta_2}{\beta_1}\right)\right]\right\}\right)
+\frac{1}{2M^6}\frac{\hat m_s^2}
{[\frac12\ln (M^2/\Lambda^2)]^{-2\gamma_1/\beta_1}}\times\nonumber\\
& &\left\{A(\mu_0)+\frac{4}{9\ln\frac{M^2}{\Lambda^2}}
\left[  B(\mu_0) + \left(
-\gamma_1\psi(3)+4\frac{\beta_2}{\beta_1^2}\ln\ln\frac{M^2}{\Lambda^2}
-\frac{4}{\beta_1\gamma_1}\left(\gamma_2-\gamma_1\frac{\beta_2}{\beta_1}\right)
\right)A(\mu_0)\right]\right\}\nonumber\\
&-&\frac{3}{7\pi^2}\frac{1}{2M^6}
\frac{\hat m_s^6}{[\frac12\ln (M^2/\Lambda^2)]^{-6\gamma_1/\beta_1}}
\left\{\frac{155}{24}-\frac{\beta_1}{2}\ln\frac{M^2}{\Lambda^2}
-\frac{\beta_1}{2}\left(\frac{6\gamma_1}{\beta_1}+1\right)\psi(3)\right.
\nonumber\\
&+&\left.\left(12\frac{\beta_2}{\beta_1^2}+\frac{\beta_2}{\beta_1}\right)
\ln\ln\frac{M^2}{\Lambda^2}-\frac{12}{\beta_1\gamma_1}\left(\gamma_2-
\gamma_1\frac{\beta_2}{\beta_1}\right)\right\}\,,
\end{eqnarray}

where
\begin{eqnarray}
A(\mu_0)&=&2\langle m_s\bar uu\rangle_0 + \frac14\langle\frac{\alpha_s}{\pi}
G^2\rangle_0\left(
1+\frac{16}{9}\frac{\alpha_s(\mu_0)}{\pi}\right)\nonumber\\
&+&\langle m_s\bar ss\rangle_0
\left(1-\frac{4}{9}\frac{\alpha_s(\mu_0)}{\pi}\right)
+m_s^4(\mu_0)\frac{3}{7\pi^2}\frac{\pi}{\alpha_s(\mu_0)}
\left( 1-
\frac{173}{72}\frac{\alpha_s(\mu_0)}{\pi}\right)\,,
\end{eqnarray}
\begin{eqnarray}
B(\mu_0)&=&\frac{46}{3}\langle m_s\bar uu\rangle_0
+\frac{121}{72} \langle\frac{\alpha_s}{\pi}
G^2\rangle_0 + \frac{64}{9}\langle m_s\bar
ss\rangle_0\nonumber\\
&+&m_s^4(\mu_0)\frac{64}{21\pi^2}\frac{\pi}{\alpha_s(\mu_0)}
\left(1-\frac{519}{512}\frac{\alpha_s(\mu_0)}{\pi}\right) \,.
\end{eqnarray}

The expression (48) represents the ``theoretical'' side of the QCD
sum rule. The ``phenomenological'' side is given by the r.h.s. of
(47), with the (hadronic) spectral function Im $\psi(s)$  written as
\begin{equation}
\mbox{Im} \psi(s)= \mbox{Im} \psi(s)|_{Res}\,\, \theta(s_{0}-s) +
\mbox{Im} \psi(s)|_{QCD} \,\,\theta(s - s_{0})\,,
\end{equation}
where the first term above
describes the contributions of the resonances up to $s = 6.8 GeV^{2}$
and the second term, i.e. the hadronic
continuum, is identified as usual with the perturbative QCD expression,
which in this case is given by
\begin{equation}
\frac{1}{\pi}\mbox{Im}\,\psi(s)|_{QCD}=\frac{3}{8\pi^2}m_s^2(s)s
\left(1+\frac{17\alpha_s(s)}{3\pi}\right)\,.
\end{equation}
The continuum threshold is expected to be close to the upper
limit of the experimental data, i.e. $s_{0} \simeq 6 - 7 \mbox{GeV}^{2}$.
In principle, though, $s_{0}$ is a free parameter. Predictions will be
meaningful provided they do not depend strongly on the value of this
parameter.\\

Chiral dynamics provides a strong constraint on the behaviour of the
hadronic spectral function near threshold, viz.
\begin{equation}
\frac{1}{\pi}\mbox{Im}\,\psi(s) = \frac{3}{32\pi^2}
|d(s_{+})|^2
\sqrt{(1-\frac{s_+^{K\pi}}{s})(1-\frac{s_-^{K\pi}}{s})}\,,
\end{equation}
with $s_\pm^{K\pi}=(M_K\pm M_\pi)^2$, and $|d(s_{+})| \simeq 0.3 \mbox{
GeV}^2$.
A good fit to the
experimental data \cite{5} is obtained by using (53), which simulates
non-resonant background, to normalize two Breit-Wigner  forms
for the $K_{0}^{*} (1430)$ and $K_{0}^{*} (1950)$ resonances, with
masses and widths:  $M_{1} = (1.40 \pm 0.01)$ GeV, $\Gamma_{1} =
(325\pm 30)$ MeV,
$M_{2} = (1.94\pm 0.03)$ GeV, $\Gamma_{2} = (450\pm 100)$ MeV.\\

As for the QCD parameters,
we adopt the following values for the nonperturbative condensates:
$\langle \bar uu\rangle_{\mu_0}=-(0.25)^3$ GeV$^3$ at a scale
$\mu_0=1$ GeV and $\langle \bar ss\rangle_{\mu_0}/\langle \bar uu
\rangle_{\mu_0} = 0.7 - 1$. The gluon condensate has been extracted
some time ago \cite{Bert} from data on $e^{+}-e^{-}$ annihilation, and
$\tau$-decay, with  values in the range:
$\langle\frac{\alpha_s}{\pi} G^2\rangle=0.02-0.06$ GeV$^4$. The QCD scale
for three flavours is $\Lambda \simeq$  200-400 MeV \cite{PDG,Barb}.

The invariant strange quark mass $\hat m_s$ is determined by solving the
equation resulting from inserting (51) on the r.h.s. of (47)
and using (48) on the l.h.s.. Typical results are shown in
Fig.3 (for $\Lambda=200$ MeV), and Fig.4 (for $\Lambda=400$ MeV),
corresponding to
$\langle\frac{\alpha_s}{\pi} G^2\rangle_{\mu_0}=0.02$ \mbox{GeV}$^4$.
In both these figures we used
$\langle \bar ss\rangle_{\mu_0}/\langle \bar uu \rangle_{\mu_0} = 1$.
Results are essentially unchanged if this ratio deviates from unity by
some 30\%. The error
on $m_{s}$ is determined by its variation when all relevant parameters
are changed within the ranges indicated above. This gives, for the two
extreme choices of $\Lambda$,
\begin{equation}
\hat m_s = 213 - 222 \mbox{MeV}\,, \quad \bar m_s (1 \mbox{GeV})
= 171 - 179 \mbox{MeV}\,\,\,\,\, (\Lambda = 200 \mbox{MeV})\,,
\end{equation}
\begin{equation}
\hat m_s = 142 - 147 \mbox{MeV}\, ,\quad \bar m_s (1 \mbox{GeV})
= 162 - 168 \mbox{MeV}\,\,\,\,\, (\Lambda = 400 \mbox{MeV})\,,
\end{equation}
where the variation of the strange-quark mass, for a given value of
$\Lambda$, reflects the uncertainties in the values of
the gluon condensate and $s_{0}$.\\

The results of this determination show a welcomed stability in the
Borel variable $M^{2}$, as well as in the continuum threshold $s_{0}$.
To estimate the error induced by the uncertainties in the hadronic spectral
function, we have varied the resonance parameters within the limits shown
above. This gives an additional error of about $\pm 7$ MeV.
The final uncertainty in $m_{s}$ is almost exclusively due to the
influence of $\Lambda$ and the gluon condensate.

   The effect of the three-loop radiative correction to $\Pi_{0}$, and
hence to $\psi''$, has been to reduce the value of the invariant mass
$\hat m_s$ by (5-10)\%. Combining the
results in (54) and (55) into a single prediction and including the
additional error due to uncertainties in the hadronic parameters, leads to
\begin{equation}
\hat m_s = 182 \pm 45 \mbox{MeV}\,,\quad
\bar m_s (1 \mbox{GeV}) = 171 \pm 15 \mbox{MeV}.
\end{equation}

\section{Conclusions}
In this paper we have discussed in detail how to absorb mass singularities
into the vacuum expectation value of other operators of appropriate
dimension, for the case of the mass logarithms $m^4\log(m^2/Q^2)$. We have
also included the mixing with other dimension-4 operators to two-loop
order. A comparison has been made with earlier analyses of this problem
\cite{BroadGen87},\cite{CheSpi88}, \cite{T83}-\cite{BroadGen84}. In
particular, we have shown that in our approach it is possible to avoid
terms involving inverse powers of $\alpha_{s}$ which, being parametrically
enhanced, might lead to large corrections. We have then used the QCD
expression of the current correlator involving the strangeness changing
vector current, together with a fit to the experimental data
on the $I=\frac{1}{2}$, S-wave $K\pi$ amplitude, to determine the
strange-quark mass through a Borel QCD sum rule. Our results for
$m_{s}$ are in agreement, within errors, with the determination
of \cite{CePa}, which used the same fit to the data, but employed
the QCD approach of \cite{BroadGen87} to remove mass singularities.
The errors we quote for the strange-quark mass are larger than those
in \cite{CePa}. This is mostly due to the fact that in \cite{CePa} the
gluon condensate was fixed at the single value
$\langle\frac{\alpha_s}{\pi} G^2\rangle=0.03$ \mbox{GeV}$^4$, and
$\Lambda$ was allowed to change in the narrower interval
$\Lambda = 100 - 200 \mbox{MeV}$.\\

\noindent
{\bf Acknowledgements}\\
One of us (D.P.) is grateful to M.Jamin for useful discussions on the
subject of this paper and for interesting comments on the manuscript.
Many thanks are due to Gagyi-Palffy Zoltan for his help with processing the
figures.
The work of (CAD) has been supported in part by the Foundation for Research
Development, and by the John Simon Guggenheim Memorial Foundation. The
work of D.P. has been supported by Graduiertenkolleg Teilchenphysik,
Universit\"{a}t Mainz.\\

\vspace*{1cm}
\section*{Figure Captions}
\vspace*{1cm}
\begin{description}
\begin{itemize}
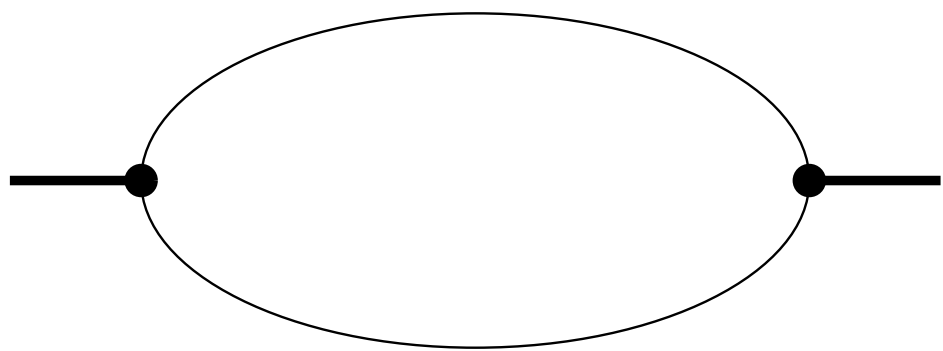
\item[Figure 1.] Lowest order contribution to the correlator $\psi(Q^2)$.
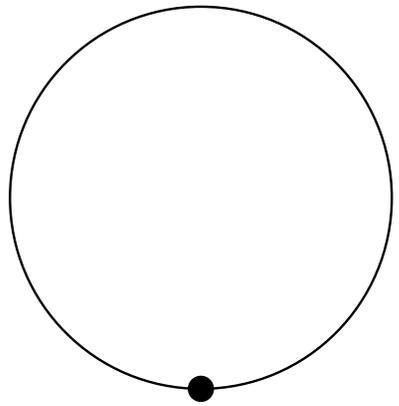
\item[Figure 2.] Vacuum diagram contributing to the perturbative VEV of the
		     operator $\bar ss$.
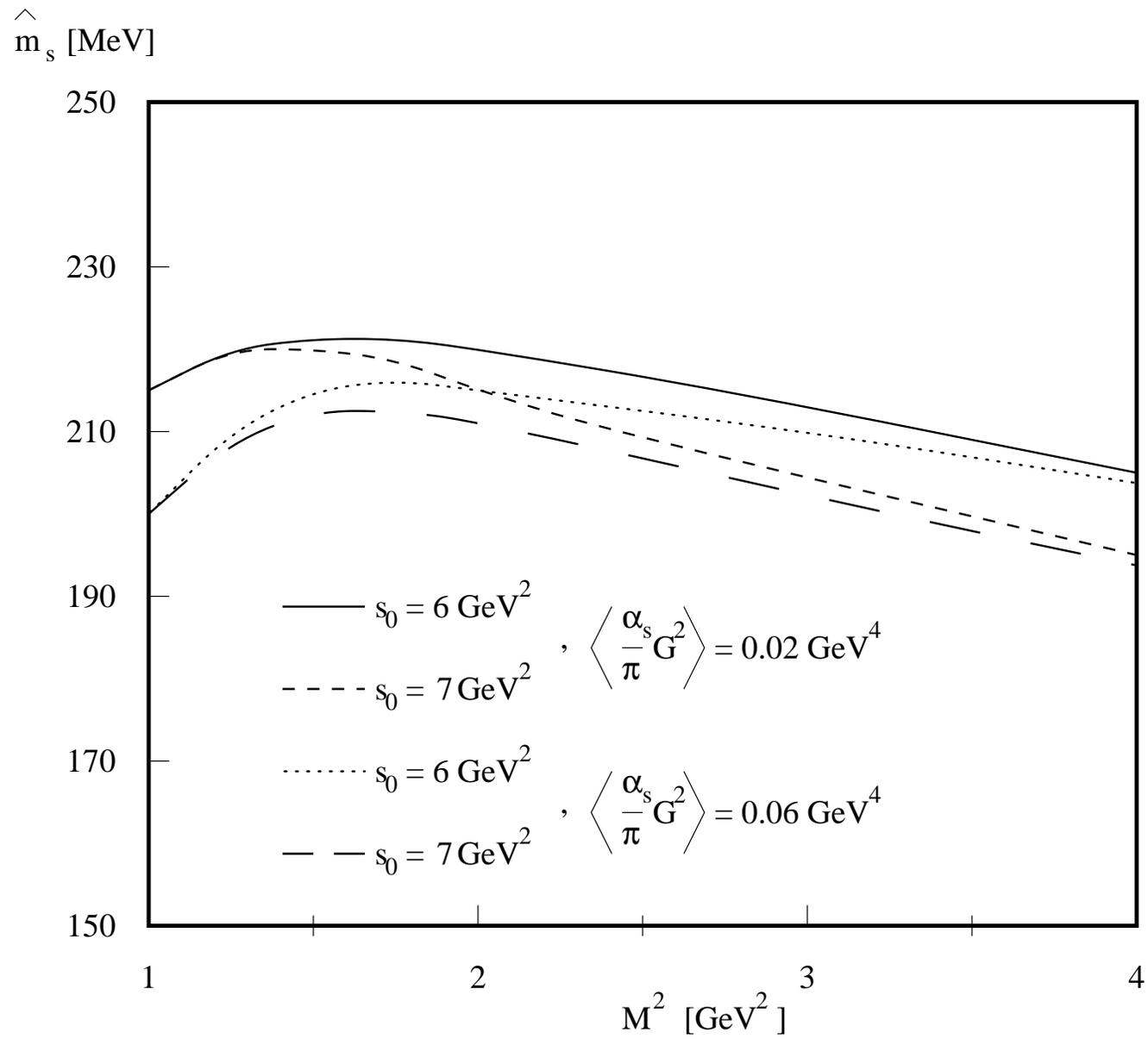
\item[Figure 3.] The invariant strange quark mass $\hat m_s$ as a function of
   the Borel variable $M^2$, for $\Lambda$ = 200 MeV. The values of the gluon
   condensate $\langle\frac{\alpha_s}{\pi} G^2\rangle_{\mu_0}$ and of the
   continuum threshold $s_0$ have been varied between 0.02 and 0.06 GeV$^4$
   and respectively, $s_0 = 6$ and 7 GeV$^2$.
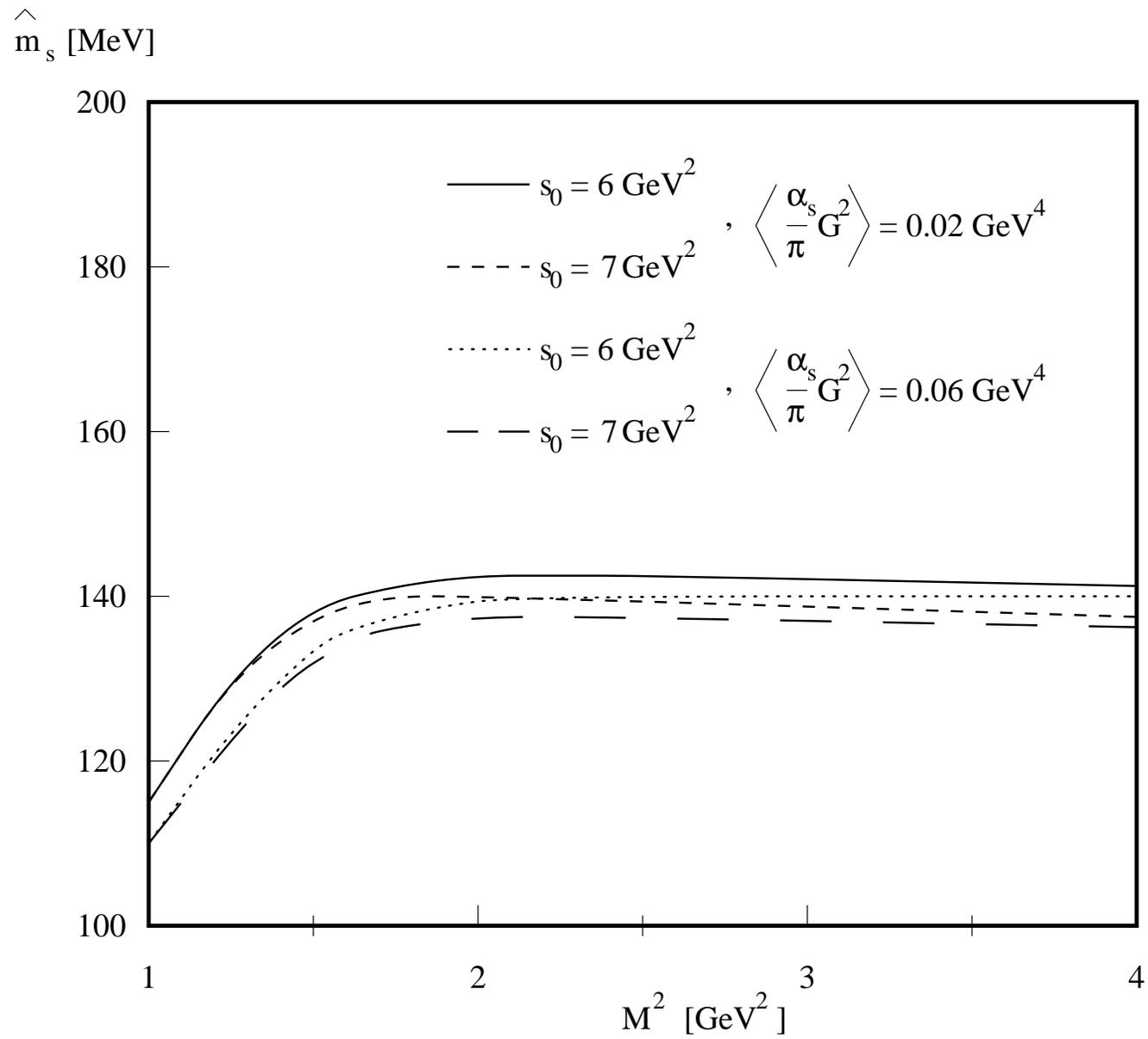
\item[Figure 4.] The same as Fig.3, except for $\Lambda$ = 400 MeV.
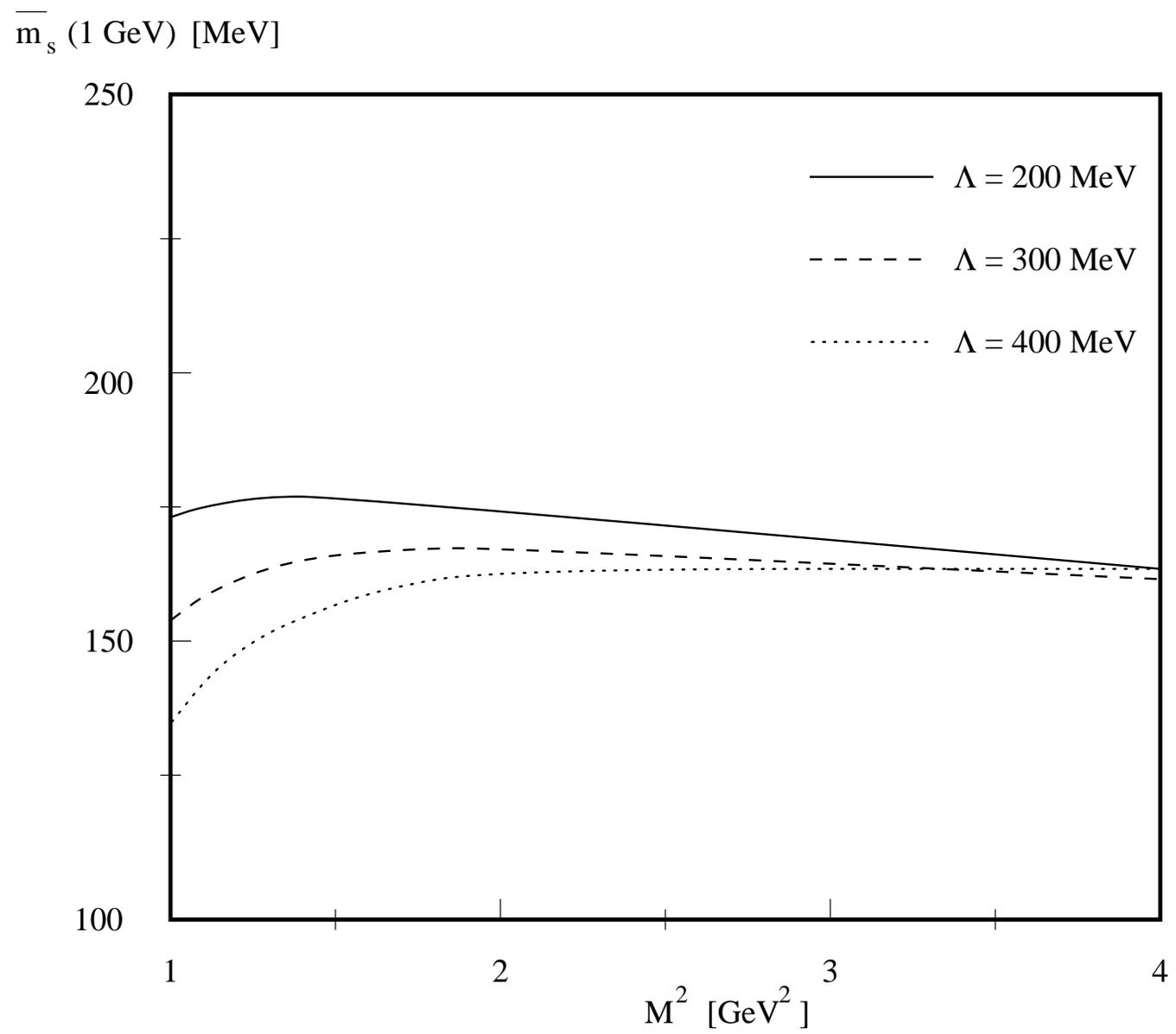
\item[Figure 5.] Results for the running strange quark mass
		     $m_s(1$ GeV) at the scale $\mu=1$ GeV.
    a) $\Lambda=200$ MeV, b) $\Lambda = 300$ MeV, c) $\Lambda = 400$ MeV.
 The continuum threshold $s_0=6.5$ GeV$^2$.
\end{itemize}
\end{description}

\begin{thebibliography}{9}
\bibitem{ft1} In particular
it also excludes the normal ordering, as the latter amounts to a
specific non-minimal subtraction of diagrams contributing to VEV's of
composite operators.
\bibitem{ft2} We would like to stress that Eq.~(31) is now a very well
   established result as it was recently confirmed by two independent
   calculations \protect\cite{Surguladze94,me94}.
\end{thebibliography}
\end{document}